\documentclass{article}
\usepackage{fullpage}
\usepackage{graphics,psfrag,epsfig}
\usepackage{amsfonts,latexsym,eucal,amsmath,amsthm,amssymb}

\begin{document}

\newcommand{\rum}{\rule{0.5pt}{0pt}}
\newcommand{\rub}{\rule{1pt}{0pt}}
\newcommand{\rim}{\rule{0.3pt}{0pt}}
\newcommand{\numtimes}{\mbox{\raisebox{1.5pt}{${\scriptscriptstyle \times}$}}}
\newcommand{\optprog}[2]
{%
  \noindent\mbox{}\\[0cm]
  \noindent\fbox{%
  \begin{minipage}{0.955\linewidth}
    \mbox{}\\[-0.5cm]
    #1\\[#2]
  \end{minipage}
  }
  \noindent\mbox{}\\[-0.2cm]
}

\renewcommand{\refname}{References}

\twocolumn[%
\begin{center}
{\Large\bf Dialogue Concerning The Two Chief World Views \rule{0pt}{13pt}}\par
\bigskip
Craig Alan Feinstein \\ {\small\it 2712 Willow Glen Drive, Baltimore, Maryland
21209\rule{0pt}{13pt}}\\ \raisebox{-1pt}{\footnotesize E-mail: cafeinst@msn.com,
BS"D}\par
\bigskip\smallskip
{\small\parbox{11cm}{%
\bigskip \noindent \textbf{Abstract:} In 1632, Galileo Galilei wrote a 
book called \textit{Dialogue Concerning the Two Chief World Systems} which 
compared the new Copernican model of the universe with the old Ptolemaic model. 
His book took the form of a dialogue between three philosophers,
Salviati, a proponent of the Copernican model, Simplicio, a
proponent of the Ptolemaic model, and Sagredo, who was initially
open-minded and neutral. In this paper, I am going to use Galileo's
idea to present a dialogue between three modern philosophers, Mr.
Spock, a proponent of the view that $\mathsf{P} \neq \mathsf{NP}$,
Professor Simpson, a proponent of the view that
$\mathsf{P} = \mathsf{NP}$, and Judge Wapner, who
is initially open-minded and neutral.

\bigskip \noindent \textbf{Disclaimer:} This article was authored
by Craig Alan Feinstein in his private capacity. No official support or endorsement by
the U.S. Government is intended or should be inferred.\rule[0pt]{0pt}{0pt}}}
\bigskip
\end{center}]{%

Since 2006, I have published four proofs that $\mathsf{P} \neq
\mathsf{NP}$ \cite{b:CF06,b:CF11,b:CF11b,b:CF12}. Yet at the present
time, if one asks the average mathematician or computer scientist
the status of the famous $\mathsf{P}$ versus $\mathsf{NP}$ problem,
he or she will say that it is still open. In my opinion, the main
reason for this is because most people, whether they realize it or
not, believe in their hearts that $\mathsf{P} = \mathsf{NP}$, since
this statement essentially means that problems which are easy to
state and have solutions which are easy to verify must also be easy
to solve. For instance, as a professional magician, I have observed
that most laymen who are baffled by an illusion are usually
convinced that the secret to the illusion either involves
extraordinary dexterity or high technology, when in fact magicians
are usually no more dexterous than the average layman and the
secrets to illusions are almost always very simple and low-tech; as
the famous designer of illusions, Jim Steinmeyer, said, ``Magicians
guard an empty safe"\cite{b:S04}. The thinking that extraordinary
dexterity or high technology is involved in a magician's secret is,
in my opinion, due to a subconscious belief that $\mathsf{P} =
\mathsf{NP}$, that problems which are difficult to solve and easy to
state, in this case ``how did the magician do it?", must have
complex solutions.

I have had many conversations in which I have tried to convince all types of
people, from Usenet trolls to graduate students to professors to
famous world-class mathematicians, that $\mathsf{P} \neq
\mathsf{NP}$ with very little success; however, I predict that there
will soon come a day when the mainstream mathematics and computer
science community will consider people who believe that $\mathsf{P}
= \mathsf{NP}$ to be in the same league as those who believe it is
possible to trisect an angle with only a straightedge and compass
(which has been proven to be impossible) \cite{b:We}.

I got the idea to write this paper after I learned of Galileo's book
\textit{Dialogue Concerning The Two Chief World Systems}\cite{b:GG},
which presents a dialogue between three philosophers, Salviati, a
proponent of the new Copernican model, Simplicio, a proponent of the
old Ptolemaic model, and Sagredo, who was initially open-minded and
neutral. The dialogue that follows is a dialogue between three
modern philosophers, Mr. Spock, a proponent of the view that
$\mathsf{P} \neq \mathsf{NP}$, Professor Simpson, a proponent of the
view that $\mathsf{P} = \mathsf{NP}$, and Judge Wapner, who is
initially open-minded and neutral. Professor Simpson, who is a
fictitious anglicized straw man character like Simplicio, is a composite of
many of the people whom I have had discussions with over the years
about the $\mathsf{P}$ versus $\mathsf{NP}$ problem. He presents
many challenges and questions, all of which have been raised before
by real people, that Mr. Spock, the epitome of
truth and logic, attempts to answer. And Judge Wapner, the epitome
of open-mindedness and fairness, always listens to both sides of
their arguments before drawing conclusions.

\bigskip\noindent \textbf{Spock:} Yesterday we discussed the
$\mathsf{P}$ versus $\mathsf{NP}$ problem \cite{b:BC94,b:CLR90} and
agreed that it is a problem of not only great philosophical
importance, but also it has practical implications. We decided to
look at a proof that $\mathsf{P} \neq \mathsf{NP}$ offered by Craig
Alan Feinstein in a letter entitled ``A more elegant argument that
$\mathsf{P} \neq \mathsf{NP}$"\cite{b:CF12}. The proof is
surprisingly short and simple:

\bigskip\noindent \textit{Proof:} Consider the following problem: Let
$\{s_1,\dots,s_n\}$ be a set of $n$ integers and $t$ be another
integer. Suppose we want to determine whether there exists a subset
of $\{s_1,\dots,s_n\}$ such that the sum of its elements equals $t$,
where the sum of the elements of the empty set is considered to be
zero. This famous problem is known as the SUBSET-SUM problem.

Let $k \in \{1,\dots,n\}$. Then the SUBSET-SUM problem is equivalent
to the problem of determining whether there exist sets $I^+
\subseteq \{1,\dots,k\}$ and $I^- \subseteq \{k+1,\dots,n\}$ such
that
$$
\sum_{i \in I^+} s_i = t - \sum_{i \in I^-} s_i.
$$
There is nothing that can be done to make this equation simpler.
Then since there are $2^k$ possible expressions on the left-hand
side of this equation and $2^{n-k}$ possible expressions on the
right-hand side of this equation, we can find a lower-bound for the
worst-case running-time of an algorithm that solves the SUBSET-SUM
problem by minimizing $2^k+2^{n-k}$ subject to $k \in
\{1,\dots,n\}$.

When we do this, we find that $2^k+2^{n-k}=2^{\lfloor n/2 \rfloor}+
2^{n-\lfloor n/2 \rfloor}=\Theta(\sqrt{2^n})$ is the solution, so it
is impossible to solve the SUBSET-SUM problem in $o(\sqrt{2^n})$
time; thus, because the Meet-in-the-Middle algorithm
\cite{b:HS74,b:MvOV96,b:Woe03} achieves a running-time of
$\Theta(\sqrt{2^n})$, we can conclude that $\Theta(\sqrt{2^n})$ is a
tight lower-bound for the worst-case running-time of any
deterministic and exact algorithm which solves SUBSET-SUM. And this
conclusion implies that $\mathsf{P} \neq \mathsf{NP}$.\qed

\bigskip\noindent To me, Feinstein's proof is not only logical
but elegant too. Also, his conclusion is confirmed by history; just
as Feinstein's theorem retrodicts, no deterministic and exact algorithm that
solves SUBSET-SUM has ever been found to run faster than the
Meet-in-the-Middle algorithm, which was discovered in 1974
\cite{b:HS74,b:Woe03}.

\bigskip\noindent \textbf{Simpson:} But there is an obvious flaw in
Feinstein's ``proof": Feinstein's ``proof" only considers a very
specialized type of algorithm that works in the same way as
the Meet-in-the-Middle algorithm, except that instead of sorting two
sets of size $\Theta(\sqrt{2^n})$, it sorts one $2^k$-size set and
one $2^{n-k}$-size set. Under these restrictions, I would agree that
the Meet-in-the-Middle algorithm is the fastest deterministic and
exact algorithm that solves SUBSET-SUM, but there are still many
possible algorithms which could solve the SUBSET-SUM problem that
the ``proof" does not even consider.

\bigskip\noindent \textbf{Wapner:} Professor Simpson, where in Feinstein's proof
does he say that he is restricting the algorithms to the class of
algorithms that you mention?

\bigskip\noindent \textbf{Simpson:} He does not say so explicitly, but
it is obviously implied, since there could be algorithms that get
around his assertion that the minimum number of possible expressions
on both sides is $\Theta(\sqrt{2^n})$.

\bigskip\noindent \textbf{Spock:} How do you know that there could be
such algorithms?

\bigskip\noindent \textbf{Simpson:} I do not know, but the burden of
proof is not on me; it is on Feinstein. And he never considers these
types of algorithms.

\bigskip\noindent \textbf{Wapner:} It is true that Feinstein never explicitly
considers algorithms which work differently than the Meet-in-the-Middle algorithm,
and the burden of proof is on Feinstein to show that these types of
algorithms cannot run any faster than $\Theta(\sqrt{2^n})$ time.

\bigskip\noindent \textbf{Spock:} Professor Simpson, is the burden of proof on Feinstein to
consider in his proof algorithms which work by magic?

\bigskip\noindent \textbf{Simpson:} No, only algorithms that are realistic.

\bigskip\noindent \textbf{Spock:} Then why would you think that algorithms
that get around the assertion that the minimum total number of possible
expressions on both sides is $\Theta(\sqrt{2^n})$ are realistic?

\bigskip\noindent \textbf{Simpson:} I do not know, but the burden of proof is
not on me; it is on Feinstein.

\bigskip\noindent \textbf{Spock:} Have you considered the fact that an
algorithm which determines in $o(\sqrt{2^n})$-time whether two sets
of size $\Theta(\sqrt{2^n})$ have a nonempty intersection
\textit{must} work by magic, unless there is a way to mathematically
reduce the two sets into something simpler?

\bigskip\noindent \textbf{Wapner:} Yes, I see your point; the minimum total
number of possible expressions on each side of the SUBSET-SUM equation puts a
natural restriction on the time that an algorithm must take to solve
the SUBSET-SUM problem.

\bigskip\noindent \textbf{Simpson:} But how do you know it is impossible
to reduce the SUBSET-SUM problem into something simpler, so that the
number of possible expressions on both sides is $o(\sqrt{2^n})$?

\bigskip\noindent \textbf{Spock:} Simple algebra. Try to simplify
the SUBSET-SUM equation above. You cannot do it. The best you can do
is manipulate the equation to get $\Theta(\sqrt{2^n})$ expressions
on each side.

\bigskip\noindent \textbf{Simpson:} I'll agree that you cannot do it
algebraically, but what
about reducing the SUBSET-SUM problem to the 3-SAT problem in
polynomial-time? This can be done since 3-SAT is
$\mathsf{NP}$-complete. If there is an algorithm that can solve
3-SAT in polynomial-time, then it would also be able to solve
SUBSET-SUM in polynomial-time, contradicting Feinstein's $\Theta(\sqrt{2^n})$ lower-bound
claim.

\bigskip\noindent \textbf{Spock:} But this is magical thinking. If a
problem is shown to be impossible to solve in polynomial time, then
reducing the problem to another problem in polynomial-time will not
change the fact that it is impossible to solve the first problem in
polynomial time; it will only imply that the second problem cannot
be solved in polynomial time.

\bigskip\noindent \textbf{Wapner:} Spock is right about this. Do you have any
other objections to Feinstein's argument?

\bigskip\noindent \textbf{Simpson:} I have many objections. For instance,
Feinstein's argument can be applied when the magnitudes of the integers in the set
$\{s_1,\dots,s_n\}$ and also $t$ are assumed to be bounded by a
polynomial to ``prove" that it is impossible to solve this modified
problem in polynomial-time. But it is well-known that one can solve
this modified problem in polynomial-time.

\bigskip\noindent \textbf{Spock:} But Feinstein's argument in fact \textit{cannot} be
applied in such a circumstance, because there would only be a
polynomial number of possible values on each side of the equation,
even though the total number of possible expressions on each side is
exponential. Feinstein's argument implicitly uses the fact that the
total number of possible values on each side of the SUBSET-SUM
equation is usually of the same order as the total number of
possible expressions on each side, when there is no restriction on
the magnitude of the integers in the set $\{s_1,\dots,s_n\}$ and
also $t$.

\bigskip\noindent \textbf{Simpson:} Then here is a better objection:
Suppose the set $\{s_1,\dots,s_n\}$ and also $t$ consist of vectors
in $\mathbb Z_2^m$ for some positive integer $m$, instead of
integers. Then one could use the same argument that Feinstein uses
to ``prove" that it is impossible to determine in polynomial-time
whether this modified SUBSET-SUM equation has a solution, when in
fact one can use Gaussian elimination to determine this information
in polynomial-time.

\bigskip\noindent \textbf{Spock:} Feinstein's argument would not
apply to this situation precisely because one can reduce the equation
$$
\sum_{i \in I^+} s_i = t - \sum_{i \in I^-} s_i.
$$
to a simpler set of equations through Gaussian
elimination. But when the set $\{s_1,\dots,s_n\}$ and also $t$
consist of integers, nothing can be done to
make the above equation simpler, so Feinstein's argument is applicable.

\bigskip\noindent \textbf{Simpson:} OK, then how would you answer
this? Consider the Diophantine equation:
$$
s_1 x_1 + \dots + s_n x_n=t,
$$
where $x_i$ is an unknown integer, for $i=1,\dots,n$. One could use the same
argument that Feinstein uses to ``prove" that it is impossible to
determine in polynomial-time whether this equation has a solution,
when in fact one can use the Euclidean algorithm to determine this
information in polynomial-time.

\bigskip\noindent \textbf{Spock:} But again Feinstein's argument
would not apply to this Diophantine equation, precisely because this
Diophantine equation can be reduced via the Euclidean algorithm to the equation,
$$
\gcd(s_1,\dots,s_n)\cdot z=t,
$$
where $z$ is an unknown integer. And it is easy to determine in
polynomial-time whether this equation has an integer solution by simply
testing whether $t$ is divisible by $\gcd(s_1,\dots,s_n)$. No such reduction
is possible with the SUBSET-SUM equation.

\bigskip\noindent \textbf{Simpson:} The Euclidean algorithm is a clever
trick that has been known
since ancient times. But how do I know that another clever
trick cannot be found to reduce the SUBSET-SUM equation to something
simpler? Like for instance, if I take the greatest common
denominator of any subset of $\{s_1,\dots,s_n\}$ and it does not
divide $t$, then I can automatically rule out many solutions to
SUBSET-SUM, all at once.

\bigskip\noindent \textbf{Spock:} But such a clever trick does not always work;
what if the gcd \textit{does} divide $t$? The $\mathsf{P}$ versus
$\mathsf{NP}$ problem is a problem about the \textit{worst-case}
running-time of an algorithm, not whether there are clever tricks
that can be used to speed up the running-time of an algorithm in
some instances. Feinstein's proof only considers the \textit{worst-case}
running-time of algorithms which solve SUBSET-SUM.

\bigskip\noindent \textbf{Wapner:} Also, it is simple high school algebra that
it is impossible to make the SUBSET-SUM equation simpler than it is:
Whenever one decreases the number of possible expressions on one side of
the equation, the number of possible expressions on the
other side increases. Mathematicians can be clever, but they cannot
be clever enough to get around this fact.

\bigskip\noindent \textbf{Simpson:} OK, but what about the fact that Feinstein never
mentions in his proof the model of computation that he is
considering? To be an valid proof, this has to be mentioned.

\bigskip\noindent \textbf{Spock:} Feinstein's proof is valid in any model of
computation that is realistic enough so that the computer cannot
solve an equation with an exponential number of possible expressions
in polynomial-time, unless it is possible to reduce the
equation to something simpler.

\bigskip\noindent \textbf{Simpson:} Or what about the fact that Feinstein never
mentions in his paper the important results that one cannot prove
that $\mathsf{P} \neq \mathsf{NP}$ through an argument that
relativizes \cite{b:BGS75} or through a natural proof \cite{b:RR97}?

\bigskip\noindent \textbf{Spock:} Feinstein's proof does not relativize,
because it implicitly assumes that the algorithms that it
considers do not have access to oracles, and Feinstein's proof is
not a natural proof, since it never even deals with the circuit
complexity of boolean functions.

\bigskip\noindent \textbf{Simpson:} What about the 2010 breakthrough by
How-grave-Graham and Joux \cite{b:GJ10} which gives a probabilistic
algorithm that solves SUBSET-SUM in $o(\sqrt{2^n})$ time? I realize
that the $\mathsf{P}$ versus $\mathsf{NP}$ problem is not about
probabilistic algorithms, but what if their algorithm can be
derandomized and solved in $o(\sqrt{2^n})$ time?

\bigskip\noindent \textbf{Spock:} The algorithm by Howgrave-Graham and Joux
does not in fact solve SUBSET-SUM, because it cannot determine for
certain when there is no solution to a given instance of SUBSET-SUM;
it can only output a solution to SUBSET-SUM in $o(\sqrt{2^n})$ time
with high probability when a solution exists. Furthermore, even if
their algorithm can be derandomized, this does not guarantee that it
will run in $o(\sqrt{2^n})$ time. And Feinstein has already proven
that such a deterministic and exact algorithm is impossible.

\bigskip\noindent \textbf{Wapner:} Are there any more objections to Feinstein's
argument?

\bigskip\noindent \textbf{Simpson:} I have no more specific objections. But the
fact that the $\mathsf{P}$ versus $\mathsf{NP}$ problem has been
universally acknowledged as a problem that is very difficult to
solve and Feinstein's ``proof" is so short and simple makes it almost certain
that it is flawed. The fact that I could not give
satisfactory responses to Spock's arguments does not mean that
Feinstein is correct; Feinstein's proof has been out on the internet
for a few years now, and still the math and computer science community
as a whole does not accept it as valid. Hence, I believe that they
are right and that Feinstein is wrong.

\bigskip\noindent \textbf{Wapner:} Professor Simpson, isn't your reason
for not believing Feinstein's proof
the same reason Feinstein suggested for why most people do not believe his
proof? Because most people believe in their hearts that
$\mathsf{P}$=$\mathsf{NP}$, that problems
which are difficult to solve and easy to state, in this case the
$\mathsf{P}$ versus $\mathsf{NP}$ problem, cannot have short and
simple solutions?

\bigskip\noindent \textbf{Spock:} Indeed it is.

\bigskip\noindent \textbf{Wapner:} And yes indeed, I am convinced that
Feinstein's proof is valid and that $\mathsf{P} \neq \mathsf{NP}$.

}

\end{document}